\documentclass[%
reprint,
nofootinbib,
 amsmath,
 amssymb,
]{revtex4-1}

\usepackage{graphicx}
\usepackage{dcolumn}
\usepackage{bm}
\usepackage{amsmath}

\begin{document}

\preprint{APS/123-QED}

\title{Anomalous response of superconducting titanium nitride resonators to terahertz radiation \\}


\author{J. Bueno$^{1,*}$, P. C. J. J. Coumou$^{2}$ , G. Zheng$^{2}$, P. J. de Visser$^{1,2}$, T. M. Klapwijk$^{2,3}$, E. F. C. Driessen$^{4,5}$, S. Doyle$^{6}$, J. J. A Baselmans$^{1}$}

\affiliation{$^{1}$SRON, Netherlands Institute of Space Research, Utrecht, The Netherlands \\
$^{2}$Kavli Institute of Nanoscience, Delft University of Technology, Delft, The Netherlands \\
$^{3}$Physics Department, Moscow State Pedagogical University, 119991 Moscow, Russia \\
$^{4}$Univ. Grenoble Alpes, INAC-SPSMS, F-38000 Grenoble, France \\
$^{5}$CEA, INAC-SPSMS, F-38000 Grenoble, France \\
$^{6}$Cardiff University - Department of Physics and Astronomy, UK \\
$^{*}$ email address: j.bueno@sron.nl \\}


\date{\today}

\begin{abstract}

We present an experimental study of KIDs fabricated of atomic layer deposited TiN films, and characterized at radiation frequencies of $350$~GHz. The responsivity to radiation is measured and found to increase with increasing radiation powers, opposite to what is expected from theory and observed for hybrid niobium titanium nitride / aluminium (NbTiN/Al) and all-aluminium (all-Al) KIDs. The noise is found to be independent of the level of the radiation power. The noise equivalent power (NEP) improves with higher radiation powers, also opposite to what is observed and well understood for hybrid NbTiN/Al and all-Al KIDs. We suggest that an inhomogeneous state of these disordered superconductors should be used to explain these observations.

\end{abstract}

\pacs{Valid PACS appear here}

\maketitle

Superconducting resonators have been proposed as kinetic inductance detectors (KIDs) for sensitive multipixel radiation detection \cite{Day}. Antenna-coupled hybrid niobium titanium nitride / aluminium (NbTiN/Al) KIDs \cite{Yates, Janssen} and all-aluminium (all-Al) \cite{de Visser} have shown generation-recombination noise and photon noise limited performance. KIDs can also be constructed as lumped element kinetic inductance detectors (LEKIDs) \cite{Doyle} in which the KID is arranged as a photon absorbing area matched to free space. Aluminium LEKIDs have also shown generation-recombination and photon noise limited performance \cite{Mauskopf}. However, the low normal-state resistivity of Al makes the design of the absorber for very high frequency radiation complex. Therefore, superconductors with a high normal state resistivity have recently become of particular interest \cite{LeDuc}.

A figure of merit F to optimize the responsivity of KIDs is defined as $F = \alpha_{sc} \tau Q_{i} F_{res} / N(0) V$ \cite{LeDuc}, with $\alpha_{sc}$ the kinetic inductance fraction, $\tau$ the quasiparticle recombination time, $Q_{i}$ the internal quality factor, $F_{res}$ the resonance frequency, $N(0)$ is the single spin electron density of states at the Fermi level,  and $V$ the volume of the KID. For example, Al KIDs have a long quasiparticle recombination time (a few milliseconds) and high internal quality factors (above one million), but their kinetic inductance fraction is low and their volume large. 

Superconducting materials with a high resistivity in the normal-state are promising because of their high quality factor and a long enough relaxation time. The high normal resistance implies a large sheet inductance, resulting in a large kinetic inductance fraction, which lowers the KID volume. The high surface impedance also eases matching to free space and optimises the photon absorption. Given the high quality NbTiN resonators pioneered by Barends {\it et al.} \cite{Barends1, Barends2}, titanium nitride (TiN) has been proposed, because it has the previously mentioned properties in addition to a tuneable critical temperature, which facilitates a relatively long quasiparticle lifetime\cite{LeDuc}. Currently, several groups are studying the implementation of TiN KID devices and instruments \cite{Vissers, Sandberg, Gao, McKenney, Mazin}. However, a material like TiN has also drawn the attention of the condensed matter physics community, interested in the disorder-induced superconductor-to-insulator transition \cite{Sacepe}. It has been shown that the superconducting state becomes gradually increasingly inhomogeneous for increasing disorder, a feature absent in the traditional description of the absorption of electromagnetic radiation by a superconductor. Consequently, a careful empirical study of the direct absorption of radiation by a strongly disordered superconductor such as TiN is urgently needed.

In this letter we concentrate our experiments on atomic layer deposited (ALD) TiN exposed to $350$~GHz radiation. A number of unanticipated trends were found in these experiments: i) the responsivity to the $350$~GHz radiation \emph{increases} for increasing radiation power, opposite to what we observe in hybrid NbTiN/Al and all-Al KIDs, and what is expected for a conventional, uniform, superconducting state; ii) the optical noise equivalent power (NEP$_{opt}$) for $350$~GHz radiation is much larger than the photon noise limited NEP at low radiation powers whereas it becomes comparable at higher radiation powers; iii) the electrical NEP is much lower than the optical NEP, in contrast to what we find for the hybrid NbTiN/Al KIDs.

Several ALD TiN films have been deposited on high resistivity ($\textgreater$~$10$~k$\Omega$cm) Si ($100$) substrates covered with a thin surface layer of native silicon oxide (for details see Coumou {\it et al.} \cite{Coumou1}). The microwave properties of these TiN films have been analysed in detail by Driessen {\it et al.} \cite{Driessen} and by Coumou {\it et al.} \cite{Coumou2}. In the latter case it was compared to a study of the local values of the superconducting density of states. The microwave response deviates from standard Mattis-Bardeen theory \cite{Mattis-Bardeen}. A modified Usadel equation was used, including a disorder-dependent pair breaking parameter, which smears the BCS density of states analogous to what is known for magnetic impurities and a pair-breaking current. Physically, it refers to a ground state in which the Cooper-pairs have a finite momentum. The pair-breaking parameter modifies the quasiparticle DoS, smoothening the coherence peaks, which enter into the generalized Mattis-Bardeen expressions \cite{Driessen}. 

The local DoS of these films is independently determined using scanning tunnelling spectroscopy (STS). A clear correspondence between the STS measurements and the model is found for the least-disordered films. In contrast, the measured tunnel curves are no longer uniquely characterizing a film for the most-disordered superconducting films \cite{Coumou2}, but are found to vary laterally. For the present experiments we use a $45$~nm thick film because it has a relatively low disorder, and therefore, is expected to allow the application of a uniform model with a broadened DoS model (although it did not have the best properties as a detector). However, it is to be expected that effects of the tendency to non-uniformity of the superconductor with increasing disorder will depend on the parameters and the techniques used in the experiments. 

The $45$~nm ALD TiN film has a superconducting transition temperature $T_{c}$ of $3.2$~K, a sheet inductance $L_{s}$ of $16.5$~pH and a sheet resistance $R_{s}$ of $41.5$~$\Omega$. Two KIDs, capacitively coupled to a feedline of $\sim$~$50$~$\Omega$, were fabricated in a single layer process. The KIDs were designed to absorb $350$~GHz radiation and consist of three inductors and an interdigitated capacitor, all in parallel, forming a tank circuit with a resonance frequency $f_{res}$ of $\sim$~$3.4$~GHz. The width of the lines of the inductors is $24$~$\mu$m and the separation between adjacent lines is $60$~$\mu$m, with a total area of the inductors of $1200$~x~$986$~$\mu$m. The quality factor Q of the KIDs is $7\times10^{4}$, set by a coupling Q of $7\times10^{4}$, with an internal Q of $1.8\times10^{6}$. 

\begin{figure}
\begin{center}
\includegraphics[width=1 \linewidth, keepaspectratio]{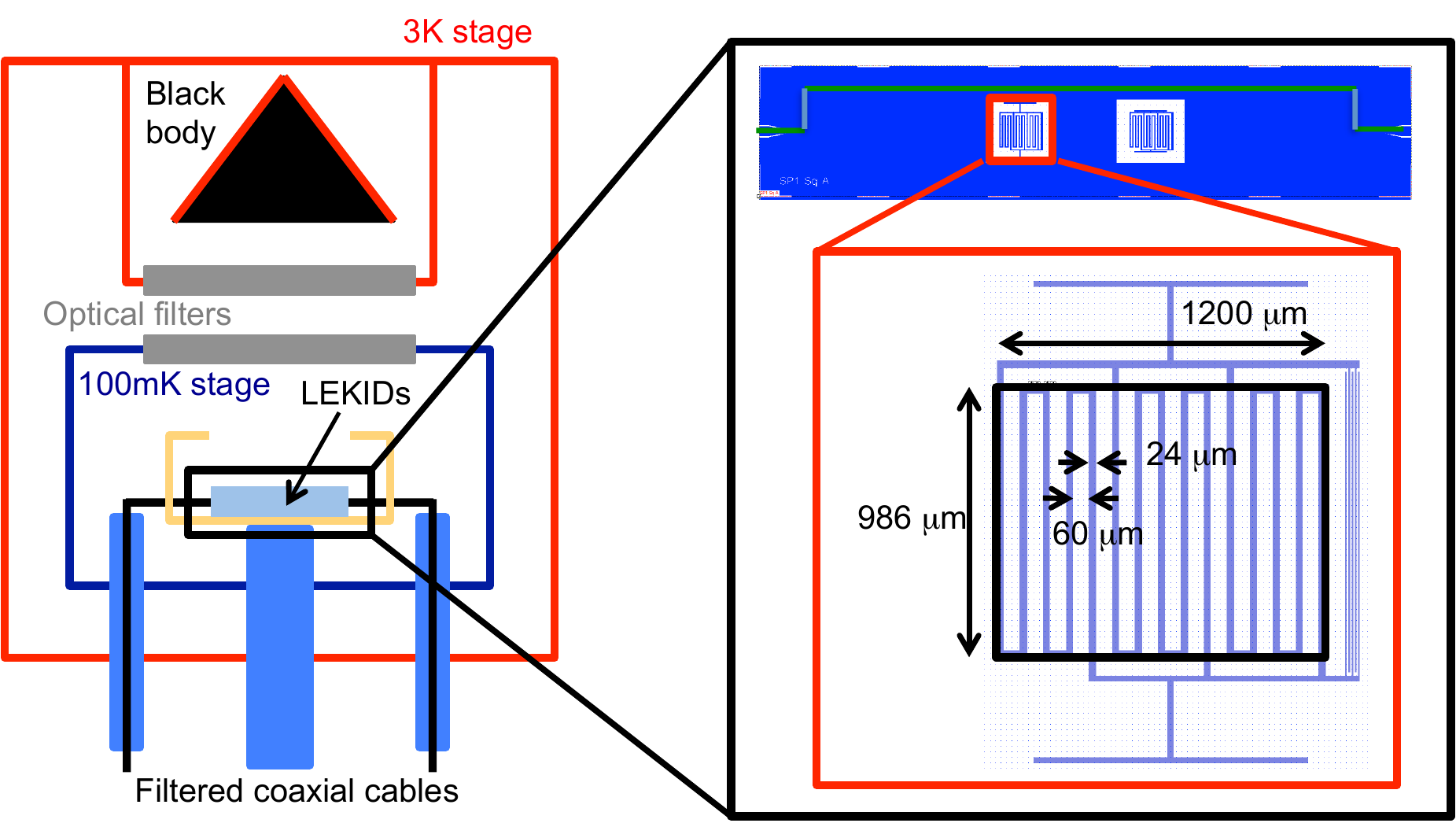}
\end{center}
\caption{\label{fig:setup}(Color online) Schematic picture of the experimental setup. {\it Left:} The KIDs are mounted in a sample box thermally anchored to the base temperature plate of the ADR ($\sim100$~mK). The sample box is surrounded by a light-tight box to prevent stray radiation from the $3$~K stage getting into the detector. A cryogenic black body source is placed at the $3$~K stage and used to illuminate the sample. The optical access to the sample box is formed by different stacks of filters placed at the black body source at the light-tight box. {\it Right:} Image of the device with the two KIDs used in these experiments, and a detailed view of one of them. The sample dimensions are $20$~x~$4$~mm, whereas each KID was roughly $1$~x~$1$~mm and designed to absorb radiation at $350$~GHz.}
\end{figure}

The experiments are performed in a pulse tube pre-cooled adiabatic demagnetisation refrigerator (ADR) with a temperature of $100$~mK, well below the superconducting transition temperature of the film. The sample is mounted in a sample holder inside a light-tight box, in a box-in-a-box configuration \cite{Baselmans1}. The $350$~GHz radiation illuminating the KIDs is coming from a cryogenic black body placed inside the cooler at the $3$~K stage, which temperature can be varied from $3$ up to $35$~K. Two stacks of metal mesh IR filters placed at the black body source and at the light-tight box determine the band pass of the radiation coupled to the KIDs, which is $50$~GHz centered around $350$~GHz. The same experimental setup is described in more detail by Janssen {\it et al.} \cite{Janssen} and by de Visser {\it et al.} \cite{de Visser}. A schematic picture of the experimental setup and device is shown in Fig. \ref{fig:setup}.

\begin{figure}
\begin{center}
\includegraphics[width=1 \linewidth, keepaspectratio]{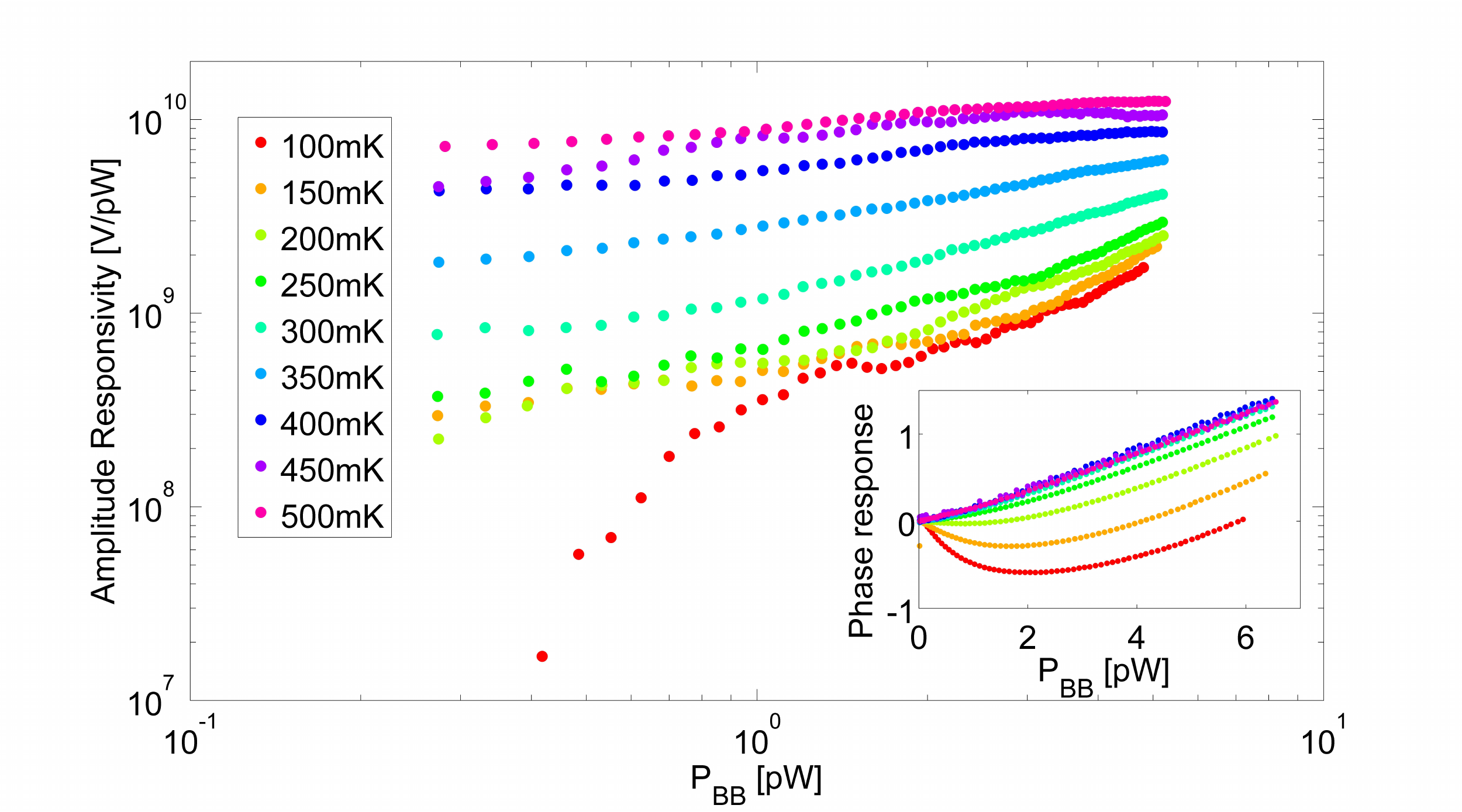}
\end{center}
\caption{\label{fig:responsivity}(Color online) Amplitude responsivity measured with a VNA as a function of radiation power at different bath temperatures. The amplitude responsivity increases when the radiation power increases, opposite to what is observed in hybrid NbTiN/Al \cite{Janssen} and all-Al \cite{de Visser} KIDs, and opposite to the usual assumptions for quasiparticle recombination. The inset shows the phase response as a function of radiation power at different bath temperatures. The phase response is presumably dominated by the TLS contribution at the lowest bath temperatures.} 
\end{figure}

The radiation power absorbed by the KIDs $P_{opt}$ is calculated by integrating the black body spectral radiance over the solid angle illuminating the KIDs, taking into account the filter transmission. Following, the conventional approach we use the number of excess quasiparticles $N_{qp}$ in a KID as the leading quantity. It is parametrized as $N_{qp} = \eta_{pb} P_{opt} \tau / \Delta$, with $\eta_{pb}$ the so-called quasiparticle creation efficiency \emph{i.e.} the number of excess quasiparticles per incoming photon, and $\Delta$ the superconducting energy gap. The absorption of the superconductor is assumed to be given by the Mattis-Bardeen theory \cite{Mattis-Bardeen} and the electron-phonon controlled recombination time as given in Kaplan {\it et al.} \cite{Kaplan}. For a photon noise limited KID, the quasiparticle lifetime at a given radiation power is expected to be inversely proportional to the square root of the absorbed power $\tau \propto P_{opt}^{-1/2}$. Therefore the $350$~GHz radiation responsivity $\delta x / \delta P_{opt}$ (where $x$ is either amplitude or phase response) is expected to be proportional to the square root of the absorbed power $\delta x / \delta P_{opt} \propto P_{opt}^{-1/2}$. These dependences were recently verified both for hybrid NbTiN/Al as well as and for the all-Al KIDs \cite{Janssen, de Visser}. A spurious contribution in the phase response (shown in the inset in Fig. \ref{fig:responsivity}), is found\footnote{At temperatures below $250$~mK, the response in the phase has an initial strong negative response followed by a gradually rising positive response, which is tentatively attributed to the temperature dependence of the dielectric material.}. Therefore, we only present and discuss the amplitude data. In the conventional response the responsivity of a KID decreases with the increase of radiation power due to the reduction in life time of quasiparticles at higher quasiparticle densities. This has been clearly shown for both hybrid NbTiN/Al \cite{Janssen} and all-Al \cite{de Visser} KIDs. In contrast, the TiN KIDs reported in this paper become more responsive at higher radiation powers, as shown in Fig. \ref{fig:responsivity}.

\begin{figure}
\begin{center}
\includegraphics[width=1 \linewidth, keepaspectratio]{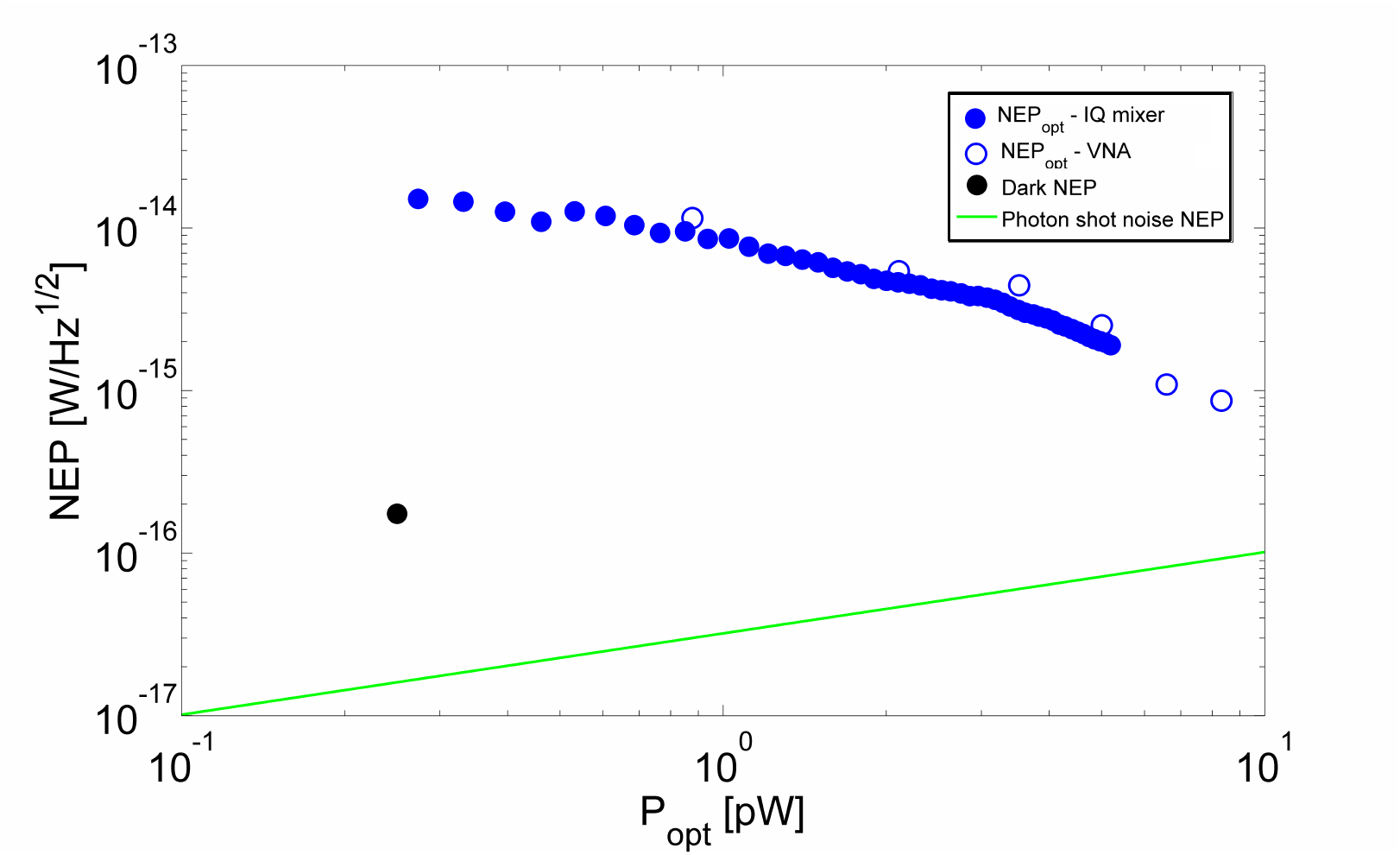}
\end{center}
\caption{\label{fig:opticalNEP}(Color online) Optical NEP for $350$~GHz radiation in the amplitude signal defined as the power spectral density divided by the responsivity measured at a modulation frequency of $100$~Hz (blue) compared to the electrical NEP calculated using the disordered-dependent pair breaking parameter model (black dot). Solid symbols were measured with a VNA whereas open symbols were measured with an IQ mixer. The green line is the expected photon noise limited NEP.} 
\end{figure}

In Fig. \ref{fig:opticalNEP}, we show the optical NEP for $350$~GHz radiation as a function of radiation power from the black-body source. The optical NEP is defined as the power spectral density divided by the responsivity $NEP_{opt}(\omega) = \sqrt{S_{A}(\omega)} / (\delta A / \delta P_{opt})$, where $S_{A}$ is the noise in the amplitude signal and $A$ the amplitude response. The noise in the amplitude signal at different radiation powers is measured at $250$~mK, still well below T$_{c}/10$, and found to be independent of radiation power. Since the noise is constant and the responsivity increases with the radiation power, we find a radiation power dependent optical NEP. Electromagnetic analytical model \cite{Blazquez} is used to evaluate the optical efficiency $\eta$ for these KIDs, showing a value of $45\%$. The measured optical NEP has been compared to the photon shot noise NEP, which is calculated from $\sqrt{2h\nu P_{opt}(1+\eta B)+2\Delta P_{opt}/\eta_{pb}} / \eta$, in which $h$ is Planck's constant, $\nu$ the radiation frequency, $P_{opt}$ the radiation power, and $(1-\eta B)$ the correction to Poissonian statistics due to photon bunching for a single mode, with $B$ the mode occupation \cite{de Visser}. The result clearly shows that the detectors are more sensitive at higher radiation powers, as is shown in Fig. \ref{fig:opticalNEP}. No sign of photon noise was observed in these experiments in contrast to recent experiments in which photon noise was reported \cite{Hubmayr}, which may be related to different photon-power densities \cite{Bradford}.

\begin{figure}
\begin{center}
\includegraphics[width=1 \linewidth, keepaspectratio]{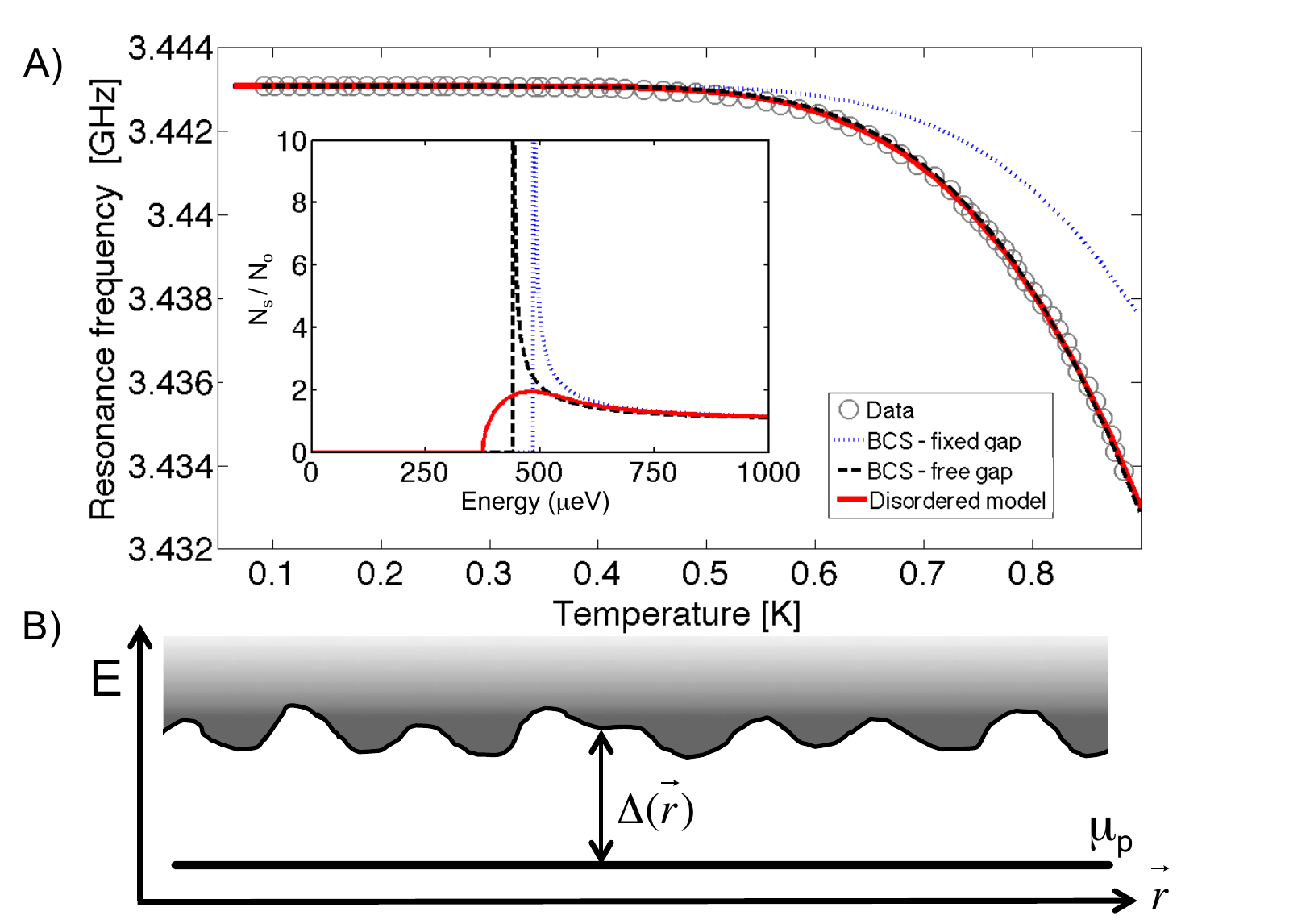}
\end{center}
\caption{\label{fig:dark}(Color online) A) Resonance frequency as a function of the bath temperature. The (grey) circles are the measured resonance frequencies, the dotted (blue) line is the prediction of the temperature dependence using Mattis-Bardeen theory \cite{Mattis-Bardeen} with the standard BCS DoS, the (black) dashed line is a fit to the Mattis-Bardeen theory with the standard BCS DoS and using the energy gap as a fit parameter, and the (red) solid line is a fit to the broadened DoS model in which both the energy gap and the broadening of the DoS are fitting parameters. The inset shows the quasiparticle DoS used for each fit. B) Cartoon of the disorder dependent landscaped energy gap conjecture. The quasiparticles get trapped in the valleys but can escape and move freely when their thermal energy becomes comparable to the roughness of the gap.} 
\end{figure}

We also calculated the electrical NEP, determined using the standard technique described by Baselmans {\it et al.} \cite{Baselmans2}, from

\begin{equation}
	NEP^{2}(\omega) = S_{A}(\omega)\left[ \frac{\eta\eta_{pb}\tau}{\Delta} \frac{\delta A}{\delta N_{qp}} \right]^{-2} (1+\omega^{2}\tau^{2}) (1+\omega^{2}\tau_{res}^{2}) 
\label{eq:darkNEP}
\end{equation}

with $\omega$ the resonator-frequency and $\tau_{res}$ the resonator ring time. The noise and quasiparticle lifetime can be measured directly, but not the electrical responsivity $\delta A / \delta N_{qp}$. Instead, the temperature dependence of the resonator $\delta A / \delta T$  has been measured (Fig. \ref{fig:dark}) with $T$ the temperature of the KID. The electrical response has been compared to Mattis-Bardeen theory \cite{Mattis-Bardeen}. The electrical response was converted into responsivity calculating the number of thermally excited quasiparticles present at the given temperature. The number of quasiparticles was calculated using the relation shown in Eq. \ref{eq:Nqp} with a standard BCS DoS

\begin{equation}
\begin{split}
	n_{qp} = 4N(0) \int_0^\infty \! N(E) f(E) \mathrm{d}E \, \approx \\ 
	\approx 2N(0) \sqrt{2\pi k_{B}T\Delta(0)}e^{-\Delta(0)/k_{B}T},
\end{split}
\label{eq:Nqp}
\end{equation}

where $N(E)$ is the normalised quasiparticle density of states, $f(E)$ the quasiparticle energy distribution, $k_{B}$ the Boltzmann constant, and $\Delta(0)$ the superconducting energy gap at zero temperature. At this point we emphasize that for our material the correct conceptual framework and theoretical model is unclear. It is not possible to describe the temperature dependence of the resonance-frequency with standard formulas, taking the measured $T_c$ and a fixed relationship between $k_{B}T_c$ and $\Delta(0)$ of $1.76$ valid for weak-coupling superconductors. This leads to the blue dotted curve in Fig. \ref{fig:dark}, and clearly signals a deviation. One can also take the energy-gap as a free adjustable parameter. This leads to an excellent fit and consequently to a DoS represented by the black dashed curve in the inset. In the more realistic model including the disorder \cite{Driessen} which leads to the red solid line, we obtain an excellent fit as well, but with a broadened DoS, as also observed with local tunnelling spectroscopy, and shown in the inset. However, as we will argue below, such an approach is potentially misleading because it suggests that one is dealing with a uniform superconducting state.

We determine the quasiparticle lifetime by measuring the change of the amplitude signal, while equilibrium is being restored, after applying a short high power microwave pulse. The pulse decay time which we interpret as the quasiparticle lifetime is found to be $\sim40$~$\mu$sec at $250$~mK (same temperature at which the optical NEP was measured). We find that it does not follow the exponential pattern expected from Kaplan {\it et al.} \cite{Kaplan}, as reported before by Coumou {\it et al.} \cite{Coumou1}. The noise is determined with the black body source present at $3.5$~K. We find that the electrical NEP equals to $7\times10^{-17}$~W/Hz$^{1/2}$, which is in agreement with the electrical NEP reported by Leduc {\it et al.} \cite{LeDuc}, taking into account the different superconducting transition temperatures and volumes. The difference between the electrical NEP and the measured optical NEP at the lowest radiation power is a factor of $100$. Although TiN is a good absorber\footnote{Room temperature transmission measurements with a vector network analyser and cryogenic FTS measurements showed an absorption efficiency of $\sim75-90\%$ respectively.}, we cannot understand its behaviour using standard BCS theory \cite{BCS}. No better understanding is achieved when a more realistic model including the disorder \cite{Driessen} is used for the analysis.

We have calculated the electrical NEP using our current understanding of strongly disordered superconductors. However, in our analysis we continue to assume an homogeneous superconducting state, whereas it has been clearly demonstrated that at a high enough degree of disorder, the superconducting energy gap $\Delta$ varies spatially \cite{Sacepe, Coumou2}. In such an inhomogeneous system, excited quasiparticles might get trapped in low-gap puddles. In addition, the film may start to act as a random array of superconducting islands coupled by weak-link type Josephson junctions \cite{Cea}, which will have its own typical optical response.


Figure~\ref{fig:dark} B shows a cartoon of such a spatially varying energy gap $\Delta(\vec{r})$. In this figure, $\mu_p$ denotes the chemical potential, and the grey shade indicates the thermal distribution of quasiparticles. We estimate the root mean square variation of the gap $\Delta_1 = \sqrt{\langle(\Delta(\vec{r}) - \langle(\Delta(\vec{r}))\rangle)^2\rangle}$ from the broadening parameter used to describe the electrodynamic response of the resonator: $\Delta_1 \approx 44~\mu e\mathrm{V}$. At a temperature of 100~mK, most quasiparticles will be trapped in the low-gap areas, since the thermal energy $k_\mathrm{B}T = 9\mu e\mathrm{V} \ll \Delta_1$. At higher temperatures, the thermal energy and the variations of the energy gap become comparable and any (photon-induced) excess quasiparticles will be free to move. A similar effect will occur with increasing radiation power: at first excess quasiparticles will be trapped in the low-gap regions, but at increasing power, the excess quasiparticles will be free to move, and will not be sensitive to the underlying gap inhomogeneity anymore. The superconductor will increasingly behave as an electronically homogeneous material. Note that in this qualitative model, there is still a coherent superconducting condensate present, which will determine the phase response of the KID, whereas the dissipative (amplitude) response will involve a combination of mobile and trapped quasiparticles. This model might therefore also explain the difference in phase and amplitude relaxation time that was observed by Gao \emph{et al.} in TiN KIDs \cite{Gao}.

In summary, we have studied ALD TiN KIDs at a radiation frequency of $350$~GHz using the same setup as used for hybrid NbTiN/Al \cite{Janssen} and all-Al KIDs \cite{de Visser}. A radiation power dependent responsivity was found, in contrast to the well understood hybrid NbTiN/Al and all-Al KIDs. Since the noise is independent of radiation power, the optical NEP is also a strong function of the radiation power. At high radiation powers, comparable to the incident radiation in ground-based telescopes, the optical NEP approaches the requirements for ground-based instruments (recently achieved by increasing the received power per unit volume of the KIDs \cite{Hubmayr}). Despite the fact that the TiN is a good absorber, the difference between the electrical NEP and the measured optical NEP at the lowest radiation power is a factor of $100$. 

The authors thank M. Spirito for room temperature transmission measurements, P. A. R. Ade for cryogenic FTS measurements, B. Blazquez and N. Llombart for electromagnetic modelling, and R. M. J. Janssen for stimulating discussions. This work was supported as part of a collaborative project, SPACEKIDS, funded via grant 313320 provided by the European Commission under Theme SPA.$2012.2.2$-$01$ of Framework Programme $7$. This work is part of the research programme of the Foundation for Fundamental Research on Matter (FOM), which is part of the Netherlands Organisation for Scientific Research (NWO). T. M. K. also acknowledges financial support from the Ministry of Science and Education of Russia under contract No. $14$.B$25$.$31$.$0007$ and from the European Research Council Advanced grant No. $339306$ (METIQUM). E. F. C. D. was financially supported by the CEA-Eurotalents program.

\end{document}